\documentclass[11pt,twoside]{article}
\usepackage{asp2010}

\resetcounters

\begin{document}

\title{Seismology of Flaring and Dormant Active Regions}
\author{R.A.~Maurya}
\affil{Department of Physics and Astronomy, Seoul National University, Seoul 151-747, Republic of Korea.}

\begin{abstract}
We study photospheric and sub-photospheric properties of active and quiet regions observed during 11--17 February 2011 including the first X-class flare X2.2 of the solar cycle 24 which occurred in the active region NOAA 11158 on 15 February 2011. The p-mode parameters and sub-photospheric flows are computed from the ring-diagrams and inversions. We found larger frequency shifts in active regions than quiet regions. The active region NOAA 11158 shows stronger twisted sub-photospheric flows than dormant active regions. The kinetic helicity density of sub-photospheric flows of the active region NOAA 11158 shows different structure on the flare day than the pre- and post flare days.      
\end{abstract}

\section{Introduction}
\label{sec:ramajor:seism:intro}

The five-minute oscillations on the photosphere are caused by trapped acoustic waves (p-modes) in different cavities inside the solar interior \citep{Ulrich1970, Leibacher1971}. Waves with shorter wavelengths are trapped shallower beneath the photosphere and are mostly likely affected by the photospheric magnetic fields and energetic activities. They provide unique observational tool to analyse the near sub-photospheric properties of active and quiet regions. 

Active regions are magnetically strong areas on the Sun's photospheric. Some of them produce energetic transients, such as, flares and CMEs. It is expected that sub-photospheric processes play crucial role in the observed photospheric and upper atmospheric transient phenomena. Furthermore the flare productive active regions may have some distinct flows in their interiors than relatively dormant, and quiet regions. The p-modes and sub-photospheric characteristic of these regions can be studied using the ring diagrams \citep{Hill1988}. Such studies have been carried out earlier \citep[][and references therein]{Maurya2009f, Maurya2010, Maurya2010d, Maurya2011a}. They reported stronger and twisted sub-photosphere flows in the active regions than quiet regions.
 
In this paper, we analyse the temporal evolution of p-modes and sub-photospheric properties of the active region NOAA 11158 and other areas during the same time period. The active region NOAA 11158 was one of the magnetically complex regions in the rising phase of solar cycle 24 which produced the first X-class flare X2.2 on 15 February 2011 and several other flares of smaller magnitudes. We believe that the active region NOAA 11158 may have distinctly different properties than other regions. 
     
\section{The Data and Analysis}
\label{sec:ramajor:seism:DataAnalys}

The observational data, for the period 11-17 February 2011, used in this study are consists of full disk Dopplergrams obtained from the Global Oscillation Network Group (GONG), line-of-sight magnetograms from the Helioseismic and Magnetic Imager (HMI) on board Solar Dynamics Observatory (SDO). The flare activity information are taken from the solar activity reports provided by the solar monitor: \url{http://www.solarmonitor.org/}.

The p-mode parameters and the sub-photospheric flows are determined using the ring-diagrams and inversions \citep{Hill2003}. We computed the vertical component of the sub-photospheric flows using the divergence of the horizontal flows employing the continuity equation \citep[see, ][]{Komm2004}. The kinetic helicity density maps were then computed using the three components of the sub-photospheric flows.

\begin{figure}[ht]
	\centering
		\includegraphics[width=1.0\textwidth,clip=,bb=51 30 528 335]{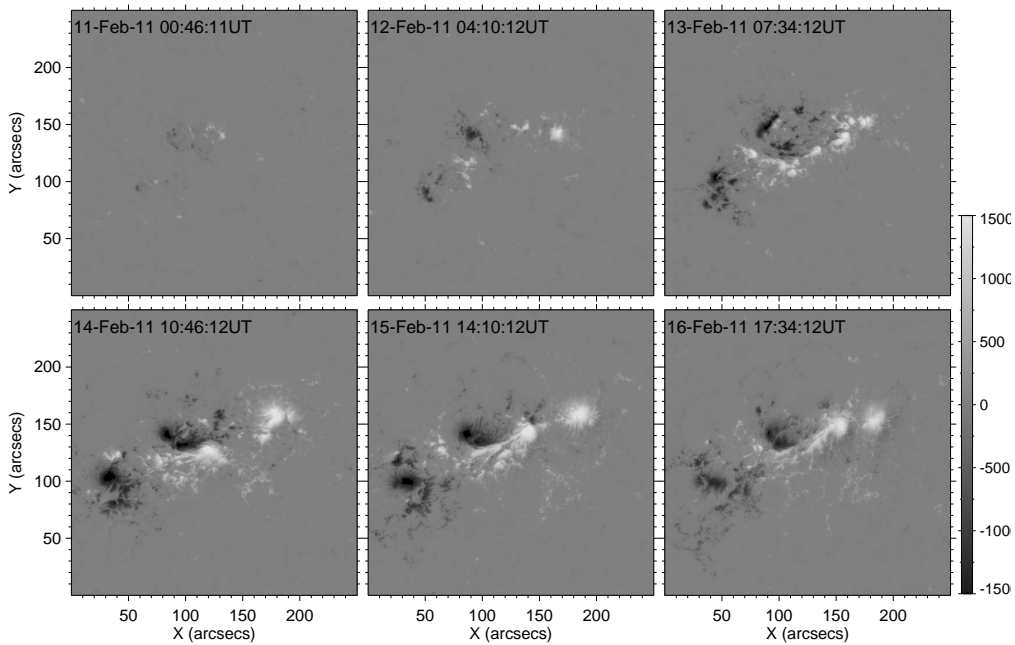}
	\caption{Temporal evolution of HMI line-of-sight magnetograms of the active region NOAA 11158 from 11 to 16 February 2011.}
	\label{fig:ramajor:seism:ar1158_evol_mag}
\end{figure}

The HMI magnetograms for different regions are extracted from the full disk images. Figure~\ref{fig:ramajor:seism:ar1158_evol_mag} shows the example of the evolution of the line-of-sight magnetic fields in the active region NOAA 11158 during 11--17 February 2011. This active region evolved very rapidly from its birth on 10 February 2011 to disappearance on the west limb on 21 February 2011. 

\begin{figure}[ht]
	\centering
		\includegraphics[width=1.0\textwidth,clip=,bb=15 10 564 370]{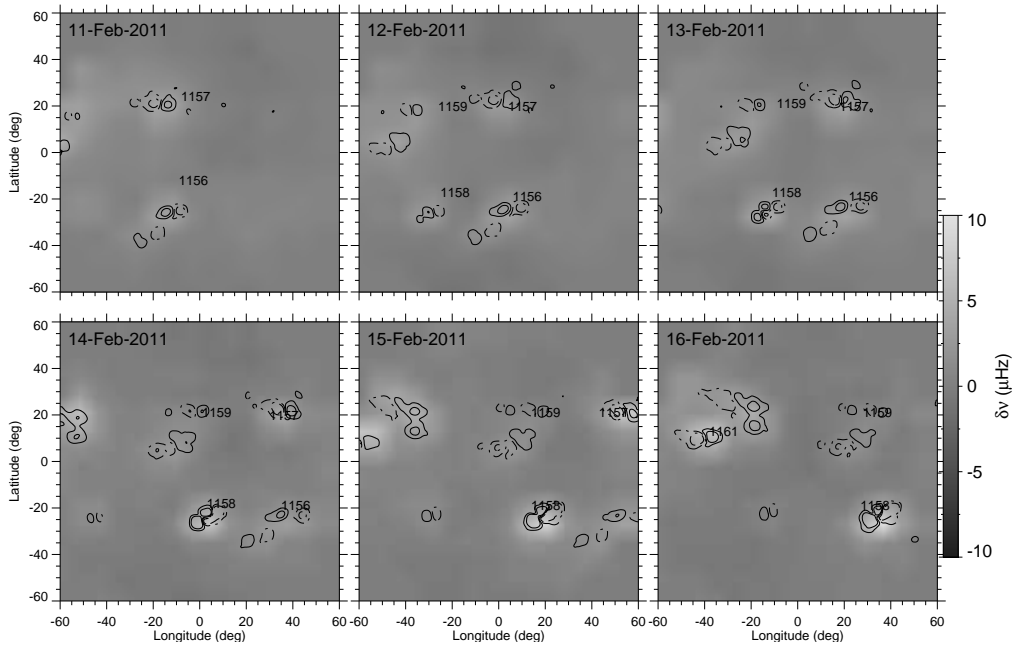}
	\caption{Maps of the p-mode frequency shift (background half tone image) during 11--16 February 2011. The overlaid contours show line-of-sight magnetic fields with labels 20 and 80\% of the negative (solid) and positive (dashed) polarities. The active regions' NOAA numbers are given at appropriate places.}
	\label{fig:ramajor:seism:pmod2107_mag}
\end{figure}

\section{Results and Discussions}
\label{sec:ramajor:seism:ResDisc}

\begin{figure}[ht]
	\centering
		\includegraphics[width=1.0\textwidth,clip=,bb=45 48 452 140]{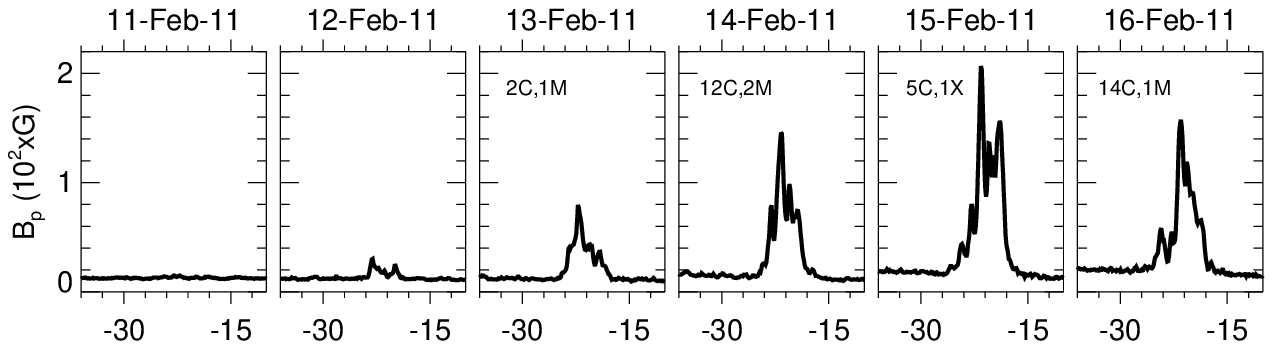}\\
		\includegraphics[width=1.0\textwidth,clip=,bb=35 11 552 307]{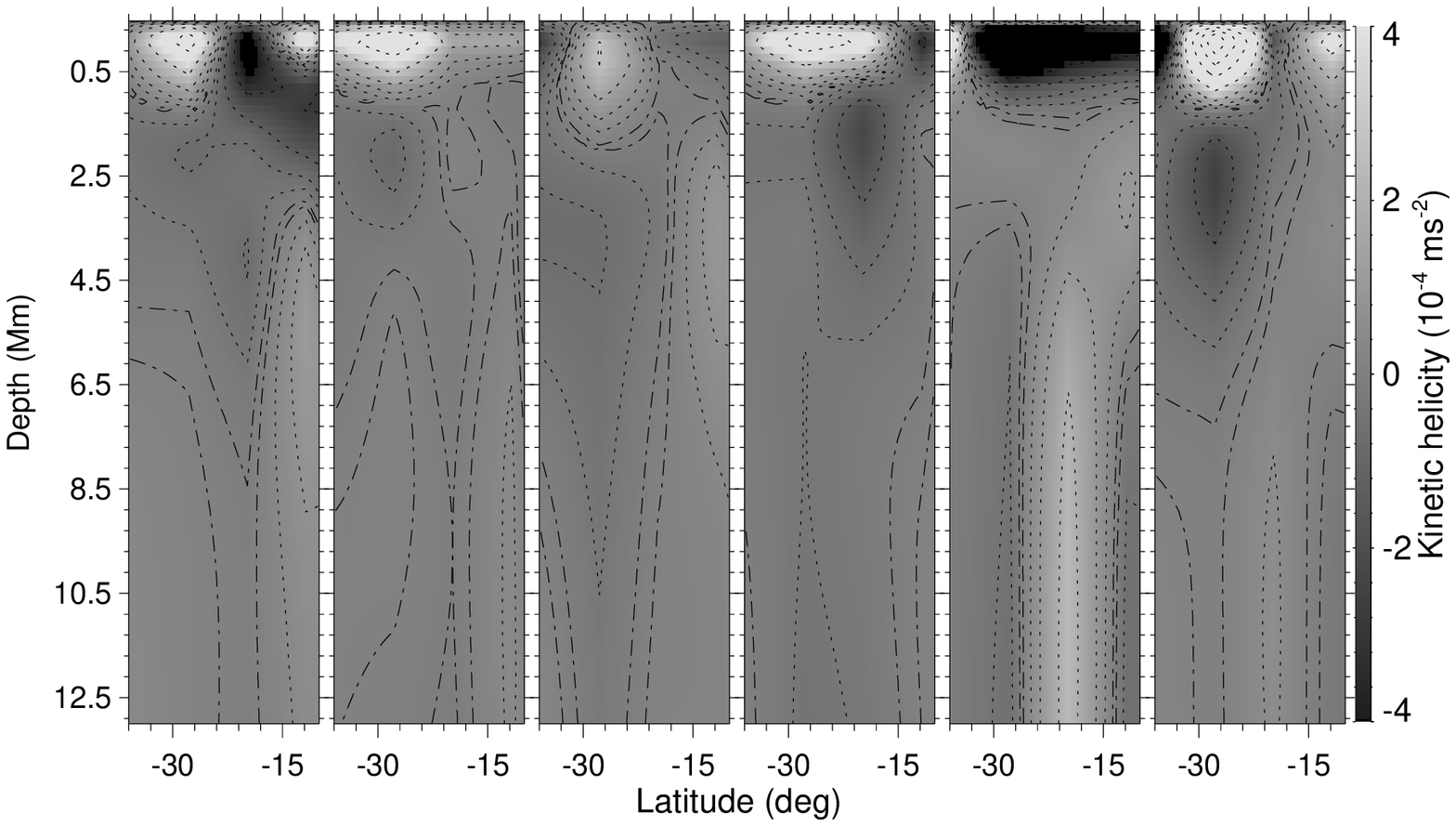}
	\caption{{\it Bottom}: The kinetic helicity density of the sub-photospheric flow at Carrington longitude of 37.5$^{\rm o}$. Different contours correspond to different levels (0.5, 2.5, 5, 10, 20, 40, 60 and 80\%) of the absolute maximum of the kinetic helicity density. {\it Top}: The averaged profiles of photospheric magnetic fields over the area of $16^{\rm o}$ longitude width. Total number of C, M, and X-class flares occurred in the active region NOAA 11158 are given in top panel.}
	\label{fig:ramajor:seism:khdmap-mag_prof}
\end{figure}

The results of the analysis are shown in the Figures~\ref{fig:ramajor:seism:ar1158_evol_mag} -- \ref{fig:ramajor:seism:khdmap-mag_prof}.

Figure~\ref{fig:ramajor:seism:pmod2107_mag} shows the frequency shift maps, averaged over radial order, $n=0,\ldots, 5$, in the area of $\pm60^{\rm o}$ in longitude and latitude of the solar disk. Contour levels show the line-of-sight magnetic fields. There are larger frequency shifts in active regions than quiet regions. Similar frequency shift has also been reported for the active and quiet regions for the period 26--30 October 2003 \citep{Maurya2011a}. The frequency shift is directly associated with flow in the interior and suggest stronger flows in active regions than in quiet regions.   

Figure~\ref{fig:ramajor:seism:khdmap-mag_prof} shows the kinetic helicity density of the sub-photospheric flows in the interior of the active region NOAA 11158. These maps are constructed by taking slices at fixed Carrington longitude, $37^{\rm o}.5$, and in latitude range from $-36^{\rm o}$ to $-10^{\rm o}$. Note that near the photosphere ($<1$\,Mm) there is a large uncertainty in the kinetic helicity density because of large uncertainty in the inverted flows. This uncertainty is caused by limited number of fitted modes due to poor spatial resolution.  

On 15 February 2011, there is strong kinetic helicity density in the interior of the active region NOAA 11158 when it had produced the largest flare X2.2. Notably, similar kinetic helicity density maps were also reported earlier in the active region NOAA 10486 on 28 October 2003, during its exceptionally high flaring activity \citep{Maurya2009f, Maurya2011a}. But, the pattern of the kinetic helicity density of the active regions NOAA 11158 and NOAA 10486 are different.
 
Before the X2.2 flare day there is a negative kinetic helicity density around depth range 1--3 Mm which disappeared on the flare day and reappeared slightly below on the next day. On the flare day there is a strong positive kinetic helicity density in the depth below 4.5 Mm. Also the magnetic field $B_p$ became stronger on 15 February 2011 and reduced after the flare day (see Figure~\ref{fig:ramajor:seism:khdmap-mag_prof} (top)). 

\section{Summary and Conclusions}
\label{sec:ramajor:seism:SumConc}

We examined the p-modes and sub-photospheric flows of the active and quiet regions during the period of  11-16 February 2011. Larger frequency shifts are observed in active regions than in quiet regions, perhaps related to the stronger flows in the interiors of active regions. The daily maps of kinetic helicity density revealed development of complex flows in the interior of the active region NOAA 11158. Large kinetic helicity density is observed beneath the active region NOAA 11158 on 15 February 2011 when it  produced the X2.2 flare. However, kinetic helicity density of the active region NOAA 11158 are not as strong in the case of active region NOAA 10486 during and before the strongest flare X17.2/4B of 28 October 2003 \citep{Maurya2011a}. In order to get better near sub-photospheric properties of the active region NOAA 11158, we plan to analyse the HMI observations. 

\acknowledgements This work utilizes data obtained by the Global Oscillation Network Group (GONG) program operated by AURA, Inc. and managed by the National Solar Observatory under a cooperative agreement with the National Science Foundation, U.S.A. This work also utilizes data from the Helioseismic and Magnetic Imager on board Solar Dynamics Observatory. The ring-diagrams and inversions are computed using the GONG ring-pipeline. R.A.M. acknowledges support by the National Research Foundation of Korea (2011-0028102).

\end{document}